\documentclass[11pt,a4paper,psamsfonts,twoside,reqno]{amsart}
\usepackage[latin1]{inputenc}
\usepackage{amssymb,amsthm}
\usepackage[mathscr]{eucal}
\usepackage{mathrsfs}
\usepackage{graphicx,pstricks}
\usepackage{scalefnt}

\theoremstyle{plain}
\newtheorem{Theorem}{Theorem}[section]

\newtheorem{Corollary}[Theorem]{Corollary}

\theoremstyle{definition}

\newtheorem{Example}[Theorem]{Example}

\theoremstyle{remark}
\newtheorem{Remark}[Theorem]{Remark}

\begin{document}
\title{Wick's theorem for $q$-deformed boson operators}
\author{Toufik Mansour}
\address{Department of Mathematics, University of Haifa, Haifa 31905, Israel}
\email{toufik@math.haifa.ac.il}
\author{Matthias Schork}
\address{Alexanderstr. 76\\ 60489 Frankfurt, Germany}
\email{mschork@member.ams.org}
\author{Simone Severini}
\address{Institute for Quantum Computing and Department of Combinatorics and Optimization, University of Waterloo, N2L 3G1 Waterloo, Canada}
\email{simoseve@gmail.com}
\date{\today}
\abstract In this paper combinatorial aspects of normal ordering arbitrary words in the creation and annihilation operators of the $q$-deformed boson are discussed. In particular, it is shown how by introducing appropriate $q$-weights for the associated ``Feynman diagrams'' the normally ordered form of a general expression in the creation and annihilation operators can be written as a sum over all $q$-weighted Feynman diagrams, representing Wick's theorem in the present context.
\endabstract
\maketitle

\noindent {\small 2000 Mathematics Subject Classification: 05A10,
05A30, 05C99}\\ \noindent {\small PACS numbers 02.10.Ox}

\section{Introduction}
Since the seminal work of Katriel \cite{kat}, the combinatorial
aspects of normal ordering arbitrary words in the creation and
annihilation operators $b^{\dag}$ and $b$ of a single-mode boson
have been studied intensively and many generalizations have also
been considered, see, e.g.,
\cite{wit0,mik,kat2,kat3,blas,blas15,scho,blas25,fuji,blas3,blas35,blas4,varv,wit,ts,tsm,tm},
and the references given therein. For an important discussion and
references to the earlier literature on normal ordering noncommuting
operators, see Wilcox \cite{wilc}. For the creation and annihilation
operator $f^{\dag}$ and $f$ of a single-mode fermion the analogous
combinatorial problem does not exist due to the nilpotency of the
operators, i.e., $(f^{\dag})^2=0=f^2$. However, if one considers
instead of a single-mode fermion a multi-mode fermion (i.e., several
sets of operators $f_i,f_i^{\dag}$) then interesting combinatorial
connections to rook numbers exist (notice that the general normal
ordering problem of a single-mode boson has also straightforward
connections to rook numbers \cite{varv}). This was noted by Navon
\cite{nav} even before Katriel demonstrated that normal ordering
powers of the bosonic number operator, i.e., $(b^{\dag}b)^n$,
involves the Stirling numbers of second kind \cite{kat}. Very
recently, combinatorial aspects of multi-mode boson operators (where
the different modes interact due to a nontrivial commutation
relation) have also been investigated \cite{tm}. Starting with the
paper of Katriel and Kibler \cite{kk}, the combinatorial aspects of
normal ordering arbitrary words in the creation and annihilation
operators $c^{\dag}$ and $c$ of a single-mode $q$-boson having the
commutation relations
\begin{equation}\label{commb}
[c,c^{\dag}]_q\equiv
cc^{\dag}-qc^{\dag}c=1,\hspace{0,3cm}[c,c]=0,\hspace{0,3cm}[c^{\dag},c^{\dag}]=0
\end{equation}
have also been considered \cite{scho,varv,kk,blas2,scho2}. Recall that {\it normal ordering} is a functional representation of operator functions in which all the creation operators stand
to the left of the annihilation operators. Let an arbitrary operator function $F(c,c^{\dag})$ be
given; a function $F(c,c^{\dag})$ can be seen as a word
on the alphabet $\{c,c^{\dag}\}$. We denote by
$\mathcal{N}_q[F(c,c^{\dag})]$ the normal
ordering of the function $F(c,c^{\dag})$. Using the
commutation relations (\ref{commb}) it is clear that its
normally ordered form
$\mathcal{N}_q[F(c,c^{\dag})]=F(c,c^{\dag})$
can be written as
\begin{equation}\label{general}
\mathcal{N}_q[F(c,c^{\dag})]=F(c,c^{\dag})=\sum_{k,l}C_{k,l}(q)(c^{\dag})^kc^l
\end{equation}
for some coefficients $C_{k,l}(q)$ and the main task
consists of determining the coefficients as explicit as possible. For example, Katriel and Kibler showed \cite{kk} that normal ordering the powers of $c^{\dag}c$ involves the $q$-deformed Stirling numbers of second kind $S_q(n,k)$ in the version of Milne \cite{milne}, i.e.,
\begin{equation}\label{katkib}
\mathcal{N}_q[(c^{\dag}c)^n]=\sum_{k=0}^n S_q(n,k)(c^{\dag})^kc^k.
\end{equation}
Note that when introducing Fock space representations one has
$c^{\dag}c=[N_c]$, where we have denoted by $N_c$ the associated
number operator and by $ [x]_q=\frac{1-q^x}{1-q}=1+q+\cdots+q^{x-1}$
the $q$-deformed version of $x$ (where $x$ is a number or an
operator). To describe the normally ordered form of an arbitrary
expression is, therefore, an interesting problem. Varvak has shown
\cite{varv} that the general coefficients can be interpreted as
$q$-rook numbers. We will describe in the present paper a different
approach associated with ``$q$-weighted Feynman diagrams''. Note
that in \cite{speicher,ansh,effros} very similar results have been
shown in slightly different situations.

The structure of the paper is as follows: In Section 2 we introduce ``Feynman diagrams'' associated to our problem (following closely the terminology of \cite{effros}) and introduce $q$-weights for these Feynman diagrams. In Section 3 we state and proof Wick's theorem adapted to the present situation, i.e., describe the coefficients of (\ref{general}) in terms of $q$-weighted Feynman diagrams. In Section 4 some examples and consequences are discussed.

\section{Feynman diagrams and associated $q$-weights}
Recall that in the undeformed case (i.e., $q=1$) one has
\begin{equation}\label{wick1q1}
\mathcal{N}[F(c,c^{\dag})]=F(c,c^{\dag})=\sum_{\pi \in \mathcal{C}(F(c,c^{\dag}))} :\pi :
\end{equation}
where we have denoted by $\mathcal{C}(F(c,c^{\dag}))$ the multiset of contractions of the word $F(c,c^{\dag})$ and the double dot operation changes the order of the operators such that all creation operators precede the annihilation operators \cite{tsm}. In comparison to \cite{tsm,tm} we now switch slightly the terminology to match the one of \cite{effros} (and also \cite{speicher,ansh,biane}) which we will follow closely. Let $S=\{s_1,\ldots,s_n\}$ be a finite linearly ordered set consisting of two types of elements, i.e., there exists a ``type-map'' $\tau : S\rightarrow \{\mathscr{A},\mathscr{C}\}$ which associates to each letter $s_i$ its type, i.e., $\tau(s_i)\in \{\mathscr{A},\mathscr{C}\}$. We call elements $s_i$ with $\tau(s_i)=\mathscr{A}$ ``annihilators'' and elements $s_j$ with $\tau(s_j)=\mathscr{C}$ ``creators''. We also denote by $S^+$ (resp. $S^-$) the set of $j$ with $\tau(s_j)=\mathscr{C}$ (resp. $i$ with $\tau(s_i)=\mathscr{A}$). A {\it Feynman diagram} $\gamma$ on $S$ is a partition of $S$ into one and two-element sets, where the two-element sets have the special property that the two elements are of different type (i.e., contain exactly one creator and one annihilator) and where the element of type $\mathscr{C}$ is the one with larger index. We also regard $\gamma$ as a set of ordered pairs $\{(i_1,j_1),\ldots,(i_p,j_p)\}$ with $i_k < j_k$ and $i_k\neq i_l, j_k\neq j_l$ and $s_{i_k}\in S^-,s_{j_k}\in S^+$. We also assume with this notation that $i_1<i_2<\cdots < i_p$.

\begin{Remark} Before continuing let us draw the connection to the terminology used in \cite{tsm,tm}.
In our concrete model, the set $S$ is given by the word
$F(c,c^{\dag})$, the two types are given by $\mathscr{C}=c^{\dag}$
and $\mathscr{A}=c$ and a Feynman diagram $\gamma$ corresponds
precisely to a contraction. In fact, the two-element sets
$(i_k,j_k)$ correspond to the edges of the contraction connecting a
creator $c^{\dag}$ with a preceding annihilator $c$. Thus, the
Feynman diagram $\gamma = \{(i_1,j_1),\ldots,(i_p,j_p)\}$
corresponds to a contraction of degree $p$.
\end{Remark}

Let us introduce some further terminology following \cite{effros}. Given $S$, we call the elements of $S$ {\it vertices}. A Feynman diagram with representation $\gamma = \{(i_1,j_1),\ldots,(i_p,j_p)\}$ is said to have {\it degree} $p$ and the two-element sets $(i_k,j_k)$ are called {\it edges}. We denote the set of all Feynman diagrams on $S$ by $\mathscr{F}(S)$ and the set of Feynman diagrams  of degree $p$ by $\mathscr{F}_p(S)$. If $|S|=n$ then $\mathscr{F}_p(S)=0$ for $p>\frac{n}{2}$. Given a Feynman diagram $\gamma \in \mathscr{F}_p(S)$ with $2p \leq n$ there will be $n-2p$ unpaired indices in $\gamma$ to which we refer as {\it singletons} (in the terminology of \cite{tsm,tm} these are the vertices of degree 0). The set of singletons of $\gamma$ will be denoted by $\mathscr{S}(\gamma)$. Let us now introduce the double dot operation for a Feynman diagram $\gamma$ on $S$. Intuitively, it means that we omit all vertices contained in the two-element sets of $\gamma$ and order the remaining singletons in such a fashion that all creators precede the annihilators. More formally, let $\gamma \in \mathscr{F}_p(S)$ and assume that $\gamma$ has $r$ singletons of type $\mathscr{C}$ (resp. $s$ singletons of type $\mathscr{A}$ with $s=n-2p+r$). Then we define $:\gamma:=\mathscr{C}^r\mathscr{A}^s$. Using this terminology, we can write the undeformed case (\ref{wick1q1}) as
\begin{equation}\label{wick1q2}
\mathcal{N}[F(c,c^{\dag})]=F(c,c^{\dag})=\sum_{\gamma \in \mathscr{F}(F(c,c^{\dag}))} :\gamma :.
\end{equation}

We now introduce a $q$-{\it weight} for Feynman diagrams such
that we can write the normally ordered form of words $F(c,c^{\dag})$ in the $q$-boson operators
in a form analogous to (\ref{wick1q2}). For this we have to introduce some more terminology following \cite{effros}.
Let $\gamma=\{(i_1,j_1),\ldots,(i_p,j_p)\}$ be a Feynman diagram. We say that a pair $(i_k,j_k)$ is a {\it left crossing}
for $(i_m,j_m)$ if $i_k<i_m<j_k<j_m$ and we define $c_l(i,j)$ to be the number of such left crossings for $(i,j)$. We define $
c(\gamma):=\sum_{(i,j)\in \gamma} c_l(i,j)$ as the {\it crossing number} of $\gamma$ (Biane calls it {\it restricted crossing
number} \cite{biane}); it counts the intersections in the corresponding graph (the {\it linear representation} of $\gamma$).
We also need to count {\it degenerate crossings}. These are the triples $i<k<j$ where $k$ is not paired (i.e., a singleton)
and $(i,j) \in \gamma$. Letting $d(i,j)$ be the number of such unpaired $k$ for the edge $(i,j)$, then $d(\gamma):=\sum_{(i,j)\in \gamma} d(i,j)$ counts
the number of such triples in $\gamma$. The {\it total crossing number} of $\gamma$ is defined by
\[
tc(\gamma):=c(\gamma)+d(\gamma).
\]
It accounts for the ``interaction'' between edges (the crossings) and the ``interaction'' between singletons and
edges (covering of singletons by edges). We now need in addition a measure which accounts for the ``interaction'' between
singletons. For a singleton $s_k$ of $\gamma$ we define its {\it length} (to the right) $l_r(s_k)$ as follows: If
the singleton $s_k$ is of type $\mathscr{C}$ then $l_r(s_k)=0$; if the singleton $s_k$ is of type $\mathscr{A}$ then $l_r(s_k)$ is
given by the number of singletons of type $\mathscr{C}$ to the right of $s_k$. The {\it length} of $\gamma$ is defined to be the
sum of the lengths of all its singletons, i.e., $l(\gamma):=\sum_{s\in \mathscr{S}(\gamma)}l_r(s)$. After these lengthy
preparations
we can now define the {\it $q$-weight} of a Feynman diagram $\gamma$ to be
\begin{equation}\label{weight1}
\mathscr{W}_q(\gamma):=q^{tc(\gamma)+l(\gamma)}.
\end{equation}

\begin{Example} Let $F(c,c^{\dag})=ccc^{\dag}ccc^{\dag}c^{\dag}cc^{\dag}cc^{\dag}c^{\dag}$ and consider the Feynman diagram $\gamma=\{(1,3),(2,6),(4,9),(5,7),(8,12)\}$ of degree 5. In its linear representation in Figure \ref{falt} the vertices of type $\mathscr{A}=c$ are depicted by an empty circle, while the vertices of type $\mathscr{C}=c^{\dag}$ are depicted by a black circle. The crossing number of $\gamma$ is given by $c(\gamma)=4$ and there are two degenerate crossings, i.e., $d(\gamma)=2$, yielding the total crossing number $tc(\gamma)=6$. There is only one singleton of type $\mathscr{A}=c$ having length $l_r(s_{10})=1$, yielding the length $l(\gamma)=1$. Thus, the $q$-weight of $\gamma$ is given by $\mathscr{W}_q(\gamma)=q^{6+1}=q^7$.

\begin{figure}[h]
\begin{center}
\begin{pspicture}(0,0)(10,.7)
\setlength{\unitlength}{3mm} \linethickness{0.3pt}
\pscircle(3,0){.3}\pscircle(3.5,0){.3}\pscircle*(4,0){.3}
\pscircle(4.5,0){.3}\pscircle(5,0){.3}\pscircle*(5.5,0){.3}
\pscircle*(6,0){.3}\pscircle(6.5,0){.3}\pscircle*(7,0){.3}
\pscircle(7.5,0){.3}\pscircle*(8,0){.3}\pscircle*(8.5,0){.3}
\linethickness{0.8pt}
\qbezier(10,.3)(11.6,1.6)(13.2,.3)
\qbezier(11.6,.3)(14.9,1.6)(18.2,.3)
\qbezier(14.9,.3)(19.05,1.6)(23.2,.3)
\qbezier(16.7,.3)(18.2,1.2)(19.9,.3)
\qbezier(21.7,.3)(25,1.6)(28.3,.3)
\put(9.7,-1.3){$1$}
\put(11.4,-1.3){$2$}
\put(13.1,-1.3){$3$}
\put(14.7,-1.3){$4$}
\put(16.4,-1.3){$5$}
\put(18.1,-1.3){$6$}
\put(19.6,-1.3){$7$}
\put(21.3,-1.3){$8$}
\put(23.0,-1.3){$9$}
\put(24.5,-1.3){$10$}
\put(26.2,-1.3){$11$}
\put(27.9,-1.3){$12$}
\end{pspicture}
\end{center}
\caption{The linear representation of the Feynman diagram $\gamma$.}
\label{falt}
\end{figure}
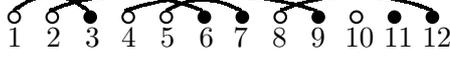
\end{Example}
\section{Wick's theorem for the $q$-deformed boson}
We can now state the generalization of (\ref{wick1q2}) to the $q$-deformed case.
\begin{Theorem}\label{qwickone} Let $F(c,c^{\dag})$ be an operator function of the annihilation and creation operators of the $q$-boson (\ref{commb}). Then the normally-ordered form $\mathcal{N}_q[F(c,c^{\dag})]$ can be described with $q$-weighted Feynman diagrams and the double dot operation as follows:
\begin{equation}\label{wick1q3}
\mathcal{N}_q[F(c,c^{\dag})]=F(c,c^{\dag})=
\sum_{\gamma \in \mathscr{F}(F(c,c^{\dag}))} \mathscr{W}_q(\gamma)\, :\gamma :.
\end{equation}
\end{Theorem}
\begin{proof}
We show this by induction in the length of the words
$F(c,c^{\dag})$. Thus, assume that the relation holds for all words
of length less than or equal $n$. A word $G(c,c^{\dag})$ of length
$n+1$ can be written either as (I) $c^{\dag}F(c,c^{\dag})$ or as (II) $cF(c,c^{\dag})$. Let us start with case (I). From the definitions it follows that
\[
\mathcal{N}_q[c^{\dag}F(c,c^{\dag})]=c^{\dag}\mathcal{N}_q[F(c,c^{\dag})]=
\sum_{\gamma \in \mathscr{F}(F(c,c^{\dag}))} \mathscr{W}_q(\gamma)\,
c^{\dag}:\gamma :=\sum_{\gamma \in
\mathscr{F}(c^{\dag}F(c,c^{\dag}))} \mathscr{W}_q(\gamma)\, :\gamma
:
\]
where we have used in the last equation that there is a bijection between $\mathscr{F}(F(c,c^{\dag}))$ and $\mathscr{F}(c^{\dag}F(c,c^{\dag}))$ such that we can identify $\gamma \in \mathscr{F}(F(c,c^{\dag}))$ with the corresponding $\gamma \in \mathscr{F}(c^{\dag}F(c,c^{\dag}))$ having the same weight. This shows case (I). Let us turn to case (II). By the induction hypothesis we assume that $\mathcal{N}_q[F(c,c^{\dag})]=\sum_{\gamma \in \mathscr{F}(F(c,c^{\dag}))}
\mathscr{W}_q(\gamma)\, :\gamma :$. Thus, from the definition of the
$q$-normal ordering we can state
\[
\mathcal{N}_q[cF(c,c^{\dag})]=\mathcal{N}_q[c\mathcal{N}_q[F(c,c^{\dag})]]=\mathcal{N}_q\left[c\sum_{\gamma \in \mathscr{F}(F(c,c^{\dag}))}
\mathscr{W}_q(\gamma)\, :\gamma :\right]
\]
which is equivalent to (write $:\gamma:=(c^{\dag})^{a_\gamma}c^{b_\gamma}$)
\[
\mathcal{N}_q[cF(c,c^{\dag})]=\sum_{\gamma \in \mathscr{F}(F(c,c^{\dag}))}
\mathscr{W}_q(\gamma)\,\mathcal{N}_q\left[c(c^{\dag})^{a_\gamma}c^{b_\gamma}\right].
\]
Using $\mathcal{N}_q[c(c^{\dag})^{a_\gamma}c^{b_\gamma}]=q^{a_\gamma}(c^{\dag})^{a_\gamma}c^{b_\gamma+1}+[a_\gamma]_q(c^{\dag})^{a_\gamma-1}c^{b_\gamma}$,
we obtain
\begin{equation}\label{part1}
\mathcal{N}_q[cF(c,c^{\dag})]=\sum_{\gamma \in \mathscr{F}(F(c,c^{\dag}))}
\mathscr{W}_q(\gamma)\left\{q^{a_\gamma}(c^{\dag})^{a_\gamma}c^{b_\gamma+1}+[a_\gamma]_q(c^{\dag})^{a_\gamma-1}c^{b_\gamma}\right\}.
\end{equation}
This is the explicit expression for the left-hand side of (\ref{wick1q3}) in the present case. We will now study the right-hand side of $(\ref{wick1q3})$ in the present case and show that it yields the same result as (\ref{part1}), thus proving the assertion for case (II). As a first step, we write $\mathscr{F}(cF(c,c^{\dag}))=\mathscr{F}^+(cF(c,c^{\dag})) \cup \mathscr{F}^-(cF(c,c^{\dag}))$ where $\mathscr{F}^+(cF(c,c^{\dag}))$ denotes the subset of Feynman diagrams where an edge starts at the most left vertex $c$ and where $\mathscr{F}^-(cF(c,c^{\dag}))$ denotes the set of the remaining Feynman diagrams where the most left vertex $c$ is a singleton. Thus,
\begin{equation}\label{basic}
\sum_{\gamma \in \mathscr{F}(cF(c,c^{\dag}))} \mathscr{W}_q(\gamma)\, :\gamma :=\sum_{\delta \in \mathscr{F}^-(cF(c,c^{\dag}))} \mathscr{W}_q(\delta)\, :\delta :+\sum_{\beta \in \mathscr{F}^+(cF(c,c^{\dag}))} \mathscr{W}_q(\beta)\, :\beta :.
\end{equation}
Note that there is a bijection between $\mathscr{F}^-(cF(c,c^{\dag}))$ and $\mathscr{F}(F(c,c^{\dag}))$ associating to $\delta \in \mathscr{F}^-(cF(c,c^{\dag}))$ the Feynman diagram $\delta'\in \mathscr{F}(F(c,c^{\dag}))$ by deleting the most left vertex $c$ in the word $cF(c,c^{\dag})$. However, the $q$-weights of the two Feynman diagrams $\delta$ and $\delta'$ are not equal. The total crossing numbers are equal, i.e., $tc(\delta)=tc(\delta')$, whereas the lengths are related by $l(\delta)=l(\delta')+a_{\delta'}$. Using \eqref{weight1}, this yields for the $q$-weights $\mathscr{W}_q(\delta)=q^{a_{\delta'}}\mathscr{W}_q(\delta')$, implying $\mathscr{W}_q(\delta):\delta:=q^{a_{\delta'}}\mathscr{W}_q(\delta'):\delta':c$. Thus, the first sum in (\ref{basic}) can be written as
\begin{equation}\label{parta}
\sum_{\delta \in \mathscr{F}^-(cF(c,c^{\dag}))}\mathscr{W}_q(\delta):\delta:=\sum_{\gamma \in \mathscr{F}(F(c,c^{\dag}))}
\mathscr{W}_q(\gamma)q^{a_\gamma}(c^{\dag})^{a_\gamma}c^{b_\gamma+1}
\end{equation}
where we have switched to the more convenient notation $\delta' \rightsquigarrow \gamma$. Let us turn to the second sum in (\ref{basic}). In analogy to before we introduce a map
\[
\mathscr{R}:\mathscr{F}^+(cF(c,c^{\dag}))\rightarrow \mathscr{F}(F(c,c^{\dag}))
\]
which (i) deletes the edge beginning at the most left vertex $c$ in the word $cF(c,c^{\dag})$, and (ii) deletes the most left vertex $c$. Clearly, this map is well-defined. In contrast to above this is not a bijection, since there can be many $\beta\in \mathscr{F}^+(cF(c,c^{\dag}))$ which are mapped onto the same $\beta'\in \mathscr{F}(F(c,c^{\dag}))$. Let us denote the preimage of $\beta'$ under this map by $\mathscr{F}^+_{\beta'}(cF(c,c^{\dag}))$, i.e.,
\[
\mathscr{F}^+_{\beta'}(cF(c,c^{\dag}))=\{\beta \in \mathscr{F}^+(cF(c,c^{\dag}))\,|\, \mathscr{R}(\beta)=\beta'\in \mathscr{F}(F(c,c^{\dag})) \}.
\]
Since these preimages are disjoint we can write
\begin{eqnarray*}
\sum_{\beta \in \mathscr{F}^+(cF(c,c^{\dag}))} \mathscr{W}_q(\beta)\, :\beta : &=&\sum_{\beta' \in \mathscr{F}(F(c,c^{\dag}))} \sum_{\beta \in \mathscr{F}^+_{\beta'}(cF(c,c^{\dag}))} \mathscr{W}_q(\beta)\, :\beta :\\
&=& \sum_{\beta' \in \mathscr{F}(F(c,c^{\dag}))} \left\{\sum_{\beta \in \mathscr{F}^+_{\beta'}(cF(c,c^{\dag}))} \mathscr{W}_q(\beta)\right\} \, (c^{\dag})^{-1}:\beta' :
\end{eqnarray*}
where we have used that for all $\beta \in
\mathscr{F}^+_{\beta'}(cF(c,c^{\dag}))$ one has
$:\beta:=(c^{\dag})^{-1}:\beta' :$. Notice that this formal notation
is not meaningless in indicating that the degree of the creation
operator has to be decreased by one, since, by definition, there is
at least one singleton of type $\mathscr{C}$ in every $\beta'$,
namely the one which becomes ``free'' after deleting the edge in
step (i) from above. Let us write as above
$:\beta':=(c^{\dag})^{a_{\beta'}}c^{b_{\beta'}}$. Assuming for the
moment
\begin{equation}\label{assum}
\sum_{\beta \in \mathscr{F}^+_{\beta'}(cF(c,c^{\dag}))} \mathscr{W}_q(\beta)=[a_{\beta'}]_q\mathscr{W}_q(\beta'),
\end{equation}
we have, therefore, shown that
\begin{equation}\label{partb}
\sum_{\beta \in \mathscr{F}^+(cF(c,c^{\dag}))} \mathscr{W}_q(\beta)\, :\beta : =\sum_{\beta' \in \mathscr{F}(F(c,c^{\dag}))} \mathscr{W}_q(\beta')[a_{\beta'}]_q \, (c^{\dag})^{a_{\beta'}-1}c^{b_{\beta'}}.
\end{equation}
Switching to the more convenient notation $\beta' \rightsquigarrow \gamma$ and
inserting (\ref{parta}) and (\ref{partb}) into the right-hand side of (\ref{basic}) yields
\begin{equation}\label{part2}
\sum_{\gamma \in \mathscr{F}(cF(c,c^{\dag}))} \mathscr{W}_q(\gamma)\, :\gamma :=\sum_{\gamma \in \mathscr{F}(F(c,c^{\dag}))}
\mathscr{W}_q(\gamma)\left\{q^{a_\gamma}(c^{\dag})^{a_\gamma}c^{b_\gamma+1}+[a_{\gamma}]_q \, (c^{\dag})^{a_{\gamma}-1}c^{b_{\gamma}}\right\}.
\end{equation}
Comparing (\ref{part1}) and (\ref{part2}) shows that $\mathcal{N}_q[cF(c,c^{\dag})]=\sum_{\gamma \in \mathscr{F}(cF(c,c^{\dag}))} \mathscr{W}_q(\gamma)\, :\gamma :$, provided (\ref{assum}) holds true. We are now going to show (\ref{assum}). First note that $|\mathscr{F}^+_{\beta'}(cF(c,c^{\dag}))|=a_{\beta'}$. The Feynman diagrams $\beta_i\in \mathscr{F}^+_{\beta'}(cF(c,c^{\dag}))$ with $1\leq i \leq a_{\beta'}$ are easy to describe: $\beta_i$ arises from $\beta'$ by adjoining the most left vertex $c$ to the word $F(c,c^{\dag})$ and connecting it via an edge with the $i$-th singleton of type $\mathscr{C}$ in the word $cF(c,c^{\dag})$. We are now going to show that
\begin{equation}\label{direct}
\mathscr{W}_q(\beta_i)=q^{i-1}\mathscr{W}_q(\beta').
\end{equation}
This implies
\[
\sum_{\beta \in \mathscr{F}^+_{\beta'}(cF(c,c^{\dag}))} \mathscr{W}_q(\beta)=\sum_{i=1}^{a_{\beta'}}\mathscr{W}_q(\beta_i)=\sum_{i=1}^{a_{\beta'}}q^{i-1}\mathscr{W}_q(\beta')=[a_{\beta'}]_q\mathscr{W}_q(\beta'),
\]
i.e., the sought-for equality (\ref{assum}). To show (\ref{direct}) we depict the word $cF(c,c^{\dag})$ in the following fashion:
\[
cF(c,c^{\dag}) \equiv v_0\,\, R_1 \,\, v_1 \,\, R_2 \,\, v_2\,\, \cdots \,\, R_{a_{\beta'}} \,\, v_{a_{\beta'}}\,\, R_{a_{\beta'}+1}
\]
where $v_0$ denotes the most left vertex $c$, the vertices $v_i$ with $1\leq i \leq a_{\beta'}$ denote the singletons of type $\mathscr{C}$ and the (possibly empty) $R_i$ the ``blocks'' in between. Let us denote by $\alpha_i$ the number of singletons of type $\mathscr{A}$ in $R_i$. For the Feynman diagram $\beta'$ we denote by $\nu_{i,k}$ the number of edges starting in $R_i$ and ending behind the vertex $v_k$. Let us now consider the Feynman diagrams $\beta_i$ and start with the case $i=1$. Due to the additional edge between $v_0$ and $v_1$ one has the following relations:
\[
c(\beta_1)=c(\beta')+\nu_{1,1}, \hspace{0,3cm} d(\beta_1)=d(\beta')+\alpha_1- \nu_{1,1},\hspace{0,3cm}l(\beta_1)=l(\beta')-\alpha_1,
\]
showing that $tc(\beta_1)+l(\beta_1)=tc(\beta')+l(\beta')$ and, therefore, $\mathscr{W}_q(\beta_1)=\mathscr{W}_q(\beta')$. Let us consider the case $i=2$. Due to the additional edge between $v_0$ and $v_2$ there will be         $\nu_{1,2}+\nu_{2,2}$ additional crossings in $\beta_2$ compared to $\beta'$, i.e., $c(\beta_2)=c(\beta')+\nu_{1,2}+\nu_{2,2}$. Turning to the degenerate crossings, there are - in comparison to $\beta'$ - three effects which have to be considered:
\begin{enumerate}
\item Since the vertex $v_2$ is no more a singleton there will be $\nu_{1,2}+\nu_{2,2}$ degenerate crossings less.
\item Since there is a new edge between $v_0$ and $v_2$ there will be $\alpha_1+\alpha_2$ degenerate crossings more, coming from the covered blocks $R_1$ and $R_2$.
\item Since $v_1$ lies in between $v_0$ and $v_2$, the triple $v_0<v_1<v_2$ also accounts for an additional degenerate crossing.
\end{enumerate}
This shows that $d(\beta_2)=d(\beta')+(\alpha_1+\alpha_2) - (\nu_{1,2}+\nu_{2,2})+1$. The length of $\beta_2$ results by decreasing the length of $\beta'$ by $\alpha_1+\alpha_2$ since the vertex $v_2$ is no more a singleton. Collecting the above results shows that $tc(\beta_2)+l(\beta_2)=tc(\beta')+l(\beta')+1$, implying $\mathscr{W}_q(\beta_2)=q\mathscr{W}_q(\beta')$. Let us now turn to the case of general $i$ (with $1\leq i \leq a_{\beta'}$) where one has the analogous ``trade off'' between the different parts which go into the $q$-weight (leaving, in effect, only the vertices $v_1,\ldots,v_{i-1}$ as contributors to the difference between $\beta_i$ and $\beta'$). The same argument as above shows that
\[
c(\beta_i)=c(\beta')+\sum_{k=1}^{i}\nu_{k,i},\hspace{0,2cm}
d(\beta_i)=d(\beta')+\sum_{k=1}^{i}\alpha_k - \sum_{k=1}^{i}\nu_{k,i}+(i-1),\hspace{0,2cm}
l(\beta_i)=l(\beta')+\sum_{k=1}^{i}\alpha_k.
\]
It follows that $tc(\beta_i)+l(\beta_i)=tc(\beta')+l(\beta')+(i-1)$ and, therefore, that $\mathscr{W}_q(\beta_i)=q^{i-1}\mathscr{W}_q(\beta')$. But this is exactly (\ref{direct}) which was to be shown. Thus, the proof for case (II) is complete.
\end{proof}
Clearly, letting $q=1$ reduces (\ref{wick1q3}) to the undeformed case (\ref{wick1q2}). Note that very similar results have been derived in \cite{speicher,ansh,effros} with a slightly different point of view. In \cite{blas35} and \cite{gough} different graphical means for normal ordering bosonic operators are discussed.

\section{Examples and consequences}
Let us consider a simple example for Theorem \ref{qwickone} before we draw some connections to $q$-rook numbers and Stirling numbers.
\begin{Example}\label{qexam} Let $F(c,c^{\dag})=c^2c^{\dag}c^2c^{\dag}=ccc^{\dag}ccc^{\dag}$. Since there are only two creators in the word, the Feynman diagrams can have degree at most two. The trivial Feynman diagram $\gamma$ of degree zero yields $:\gamma:=(c^{\dag})^2c^4$ and has $q$-weight $\mathscr{W}_q(\gamma)=q^{l(\gamma)}=q^{2+2+0+1+1+0}$. Thus, the Feynman diagram of degree zero yields the contribution $q^6(c^{\dag})^2c^4$. There are six Feynman diagrams of degree one, namely
\[
\{(1,3), (1,6), (2,3),(2,6),(4,6),(5,6) \}
\]
and their $q$-weights are given by (same order) $\{q^4,q^5,q^3,q^4,q^3,q^2\}$. Thus, the Feynman diagrams of degree one yield the contribution $(q^2+2q^3+2q^4+q^5)c^{\dag}c^3$. There are six Feynman diagrams of degree two, namely
\[
\{(1,3)(2,6),(1,3)(4,6),(1,3)(5,6),(1,6)(2,3),(2,3)(4,6),(2,3)(5,6)\}
\]
with $q$-weights $\{q^3,q^2,q,q^2,q,1\}$, yielding the contribution $(1+2q+2q^2+q^3)c^2$. Thus,
\begin{equation}\label{exqone}
\mathcal{N}_q[c^2c^{\dag}c^2c^{\dag}]=q^6(c^{\dag})^2c^4+(q^2+2q^3+2q^4+q^5)c^{\dag}c^3+(1+2q+2q^2+q^3)c^2,
\end{equation}
which may also be calculated by hand from (\ref{commb}).
\end{Example}

We now want to draw a connection between Theorem \ref{qwickone} and the results of Varvak \cite{varv}. Given a word $w=F(c,c^{\dag})$ containing $m$ creation operators $c^{\dag}$ and $n$ annihilation operators $c$ (with $n\leq m$), she associates to $w$ a certain {\it Ferrers board} $B_w$ outlined by $w$. Denoting by $R_k(B_w,q)$ the $k$th $q$-{\it rook number} of the board $B_w$, she shows that (\cite{varv}, Theorem 6.1)
\[
w=\sum_{k=0}^nR_k(B_w,q)(c^{\dag})^{m-k}c^{n-k}.
\]
Using Theorem \ref{qwickone}, we first note that the set of Feynman diagrams is the disjoint union of Feynman diagrams of degree $k$, i.e., $\mathscr{F}(w)=\cup_{k=0}^n\mathscr{F}_k(w)$, and that $\gamma \in \mathscr{F}_k(w)$ implies $:\gamma:=(c^{\dag})^{m-k}c^{n-k}$, yielding
\[
w=\sum_{k=0}^n\left\{\sum_{\gamma \in \mathscr{F}_k(w)}\mathscr{W}_q(\gamma) \right\}(c^{\dag})^{m-k}c^{n-k}.
\]
Comparing the two expressions yields the following corollary.
\begin{Corollary} Given a word $w=F(c,c^{\dag})$, the $k$th $q$-rook number of the associated Ferrers board $B_w$ equals the sum of the $q$-weights of all Feynman diagrams of degree $k$ on $w$, i.e.,
\begin{equation}
R_k(B_w,q)=\sum_{\gamma \in \mathscr{F}_k(w)}\mathscr{W}_q(\gamma).
\end{equation}
\end{Corollary}

Let us return to the undeformed case ($q=1$) and consider the particular word $F(c,c^{\dag})=(c^{\dag}c)^n$. The same argument as above shows that
\[
\mathcal{N}[(c^{\dag}c)^n]=\sum_{k=0}^{n}\sum_{\gamma \in \mathscr{F}_{n-k}((c^{\dag}c)^n)}(c^{\dag})^{k}c^{k}.
\]
Comparing this with the undeformed case of (\ref{katkib}) shows that the conventional Stirling numbers of second kind can also be interpreted as the number of Feynman diagrams of degree $n-k$ on the particular word $c^{\dag}cc^{\dag}c\cdots c^{\dag}c$ of length $2n$, i.e.,
\begin{equation}
S(n,k)=|\mathscr{F}_{n-k}((c^{\dag}c)^n)|.
\end{equation}
Turning to the $q$-deformed situation, the same argument shows that
\begin{equation}\label{qstirf}
S_q(n,k)=\sum_{\gamma\in \mathscr{F}_{n-k}((c^{\dag}c)^n)}\mathscr{W}_q(\gamma)=\sum_{\gamma\in \mathscr{F}_{n-k}((c^{\dag}c)^n)}q^{tc(\gamma)+l(\gamma)}.
\end{equation}

\begin{Remark}
In \cite{tm} the study of combinatorial aspects of the normal ordering of multi-mode boson operators was begun and interesting combinatorial questions were addressed. In view of the above $q$-Wick's theorem it is natural to consider the analogous problem for the $q$-deformed variant of the multi-mode boson operator. However, this does not seem to be easy as the following example will show. For concreteness we consider the deformed two-mode boson having the commutation relations
\[
[a,a^{\dag}]_{q_a}=1,\hspace{0,5cm}[b,b^{\dag}]_{q_b}=1,\hspace{0,5cm}[a,b^{\dag}]_{q_{ab}}=1
\]
and all other commutators vanish (note in particular that the two
modes interact nontrivially!). Here the deformation parameters
$q_a,q_b,q_{ab}$ are in the moment arbitrary. Let us consider the
simple example $F(a,a^{\dag},b,b^{\dag})=abb^{\dag}$. First
commuting $b$ and $b^{\dag}$ using the above commutation relations
yields $abb^{\dag}=q_{b}ab^{\dag}b+a$; commuting then $a$ and
$b^{\dag}$ yields $abb^{\dag}=q_bq_{ab}b^{\dag}ab+q_bb+a$. On the
other hand, first commuting $a$ and $b$ and then commuting
$b^{\dag}$ to the left yields
$abb^{\dag}=q_bq_{ab}b^{\dag}ba+q_{ab}a+b$. Clearly, for the two
results to concide we have to assume that $q_b=q_{ab}=1$. A similar
computation for $baa^{\dag}$ shows that also $q_a=1$. Thus, it seems
that does not exist a $q$-deformed version of the multi-mode boson
considered in \cite{tm}.
\end{Remark}

\end{document}